\documentclass[aps,prb,twocolumn,floatfix,superscriptaddress,amsmath,amssymb]{revtex4-2}
\usepackage{graphicx} % Required for inserting images
\usepackage[usenames,dvipsnames]{xcolor}
\usepackage{amssymb}
\usepackage{graphicx}
\usepackage{amsmath}
\usepackage{bbm}
\usepackage[colorlinks,bookmarks=true,citecolor=blue,linkcolor=red,urlcolor=blue]{hyperref}
\usepackage{braket}
\usepackage[english]{babel}
\usepackage{blindtext}
\usepackage{physics}
\usepackage{mathrsfs}
\usepackage{footmisc}
\usepackage{booktabs}
\usepackage{tikz}
\usepackage[normalem]{ulem}

\newenvironment{localgraphicspath}[1]{
  \graphicspath{#1}
}{}

\newcommand{\parR}[1]{\noindent\textbf{\textit{#1}}\hspace{0.5cm}}

\usepackage{mathtools}
\usepackage[caption=false]{subfig}

\usepackage{dsfont}

\definecolor{mypurp}{rgb}{0.35, 0, 0.7}

\usepackage{amsthm}
\usepackage{txfonts}

\usepackage{soul}

\begin{document}

%\title{Accessing Gapped Chiral Phase with Auxiliary-Assisted PEPS}
\title{Bypassing the Chiral Obstruction in Two-Dimensional Tensor Networks}

\author{Sen Niu}
\affiliation{School of Physics, Beihang University, Beijing 100191, China}

\author{Rui-Zhen Huang}
\email{huangrzh@icloud.com}
\affiliation{Graduate School of China Academy of Engineering Physics, Beijing 100193, China}

\begin{abstract}
Gapped chiral phases pose an intrinsic obstruction to projected entangled pair state (PEPS) representations with finite bond dimension $D$, which generically develop spurious long-range correlations despite the gapped nature of the target state. 
To address this long-standing problem, we take a qualitatively different route by considering the target chiral system together with a completely decoupled time-reversed auxiliary copy.
Constraining the finite-$D$ PEPS to the factorized form would simply reduce the problem to representing each chiral layer independently, leaving the original finite-$D$ obstruction unchanged.
Unexpectedly, however, we find that finite-$D$ variational optimization generates an \emph{emergent} residual interlayer entanglement, despite the absence of any explicit interlayer coupling in the Hamiltonian. This emergent entanglement is the key mechanism that bypasses the obstruction, qualitatively changing the PEPS from an effectively gapless representation with long-range correlation tails to one with a finite correlation length. We demonstrate this mechanism for both a free-fermion Chern insulator and an interacting chiral spin liquid. The residual entanglement is systematically suppressed with increasing $D$, approaching the decoupled limit, while the chiral topological information remains accessible through layer-resolved entanglement spectra. 
Our results thus uncover a previously unrecognized mechanism for bypassing the chiral obstruction through variationally generated auxiliary entanglement.
\end{abstract}

\maketitle

{\bf \emph{Introduction.}}  
Two-dimensional quantum many-body systems host a variety of unconventional phases, including topological insulators and intrinsic topological orders. Among them, gapped chiral phases form an especially important class, with examples ranging from  quantum Hall/Chern-insulator phases to chiral spin  liquids~\cite{klitzing1980new,tsui1982two,laughlin1983anomalous,haldane1988model,kalmeyer1987equivalence,nayak2008non,bartolomei2020fractional,nakamura2020direct,tang2011high,sun2011nearly,neupert2011fractional,sheng2011fractional,regnault2011fractional}.  
Chiral phases may be either short-range entangled or intrinsically topologically ordered and exhibit characteristic bulk-boundary correspondence.
Yet identifying and characterizing the topological nature of the ground state of an interacting many-body Hamiltonian remains challenging.
Tensor-network states, in particular infinite projected entangled-pair states (PEPS), provide a powerful variational framework for addressing this problem directly in the thermodynamic limit~\cite{verstraete2004renormalization,jordan2008classical,orus2014practical}.

Most non-chiral gapped states~\footnote{Here we assume translation invariance in the system.} can be efficiently represented by PEPS~\cite{gu2009tensor,buerschaper2009explicit,schuch2010peps,levin2005string, kitaev2012models,peps_bimodule}.  Chiral gapped phases, however, appear to face an intrinsic difficulty in finite-$D$ PEPS representations. 
%PEPS can reproduce many local and topological properties, but generically develop spurious power-law like long-range correlations inconsistent with the gapped target state~\cite{yang2015chiral,poilblanc2015chiral,hasik2022simulating,niu2024chiral,chen2018non,chen20203,chen2021abelian,wang2022emergent,niu2022chiral,xu2023phase,tan20241,puente2025efficient,chen2025simulating,weerda2024fractional,dong2025efficient,chen2025simulating_FCI,niu2025simulating}.
It was argued the difficulty may originate from the topological obstruction inherent to chiral topological states~\cite{wahl2013projected}, which carry nonzero (many-body) Chern numbers and cannot be Wannierized~\cite{marzari2012maximally}. 
This appears to forbid the local tensor structure of PEPS with a finite virtual bond dimension $D$. 
%This points to a fundamental obstruction to a strictly short-ranged local finite-\(D\) PEPS representatio
A no-go theorem for free-fermion Chern insulators~\cite{wahl2013projected,dubail2015tensor} establishes that Gaussian fermionic PEPS can indeed carry non-zero Chern numbers, but their correlation functions necessarily exhibit power-law decay, mimicking gapless behavior. Subsequent PEPS wavefunction construction studies found similar artifact in chiral spin liquid correlation functions~\cite{yang2015chiral,poilblanc2015chiral}. 
Increasing $D$ improves the situation only in a limited sense: Variational studies found that the physically correct short-distance exponential regime can extend to longer length scales as $D$ increases, but the long-distance tail is never eliminated; instead, it is pushed farther away and becomes the so-called ``gossamer tail''~\cite{hasik2022simulating,niu2024chiral}. This persistent fake-gapless behavior affects a broad class of PEPS simulations of chiral phases, including spin~\cite{chen2018non,chen20203,chen2021abelian,wang2022emergent,niu2022chiral,xu2023phase,tan20241,puente2025efficient,chen2025simulating}, bosonic~\cite{weerda2024fractional,dong2025efficient}, and fermionic~\cite{chen2025simulating_FCI,niu2025simulating} lattice models. Thus, despite the usefulness of PEPS in accessing chiral topological information~\cite{niu2024chiral,budaraju2024simulating}, a genuinely short-ranged finite-$D$ PEPS representation of chiral gapped physics has remained unresolved.

\begin{figure}[hbt!]
\begin{localgraphicspath}{{figs/}}
\includegraphics[width=1\columnwidth]{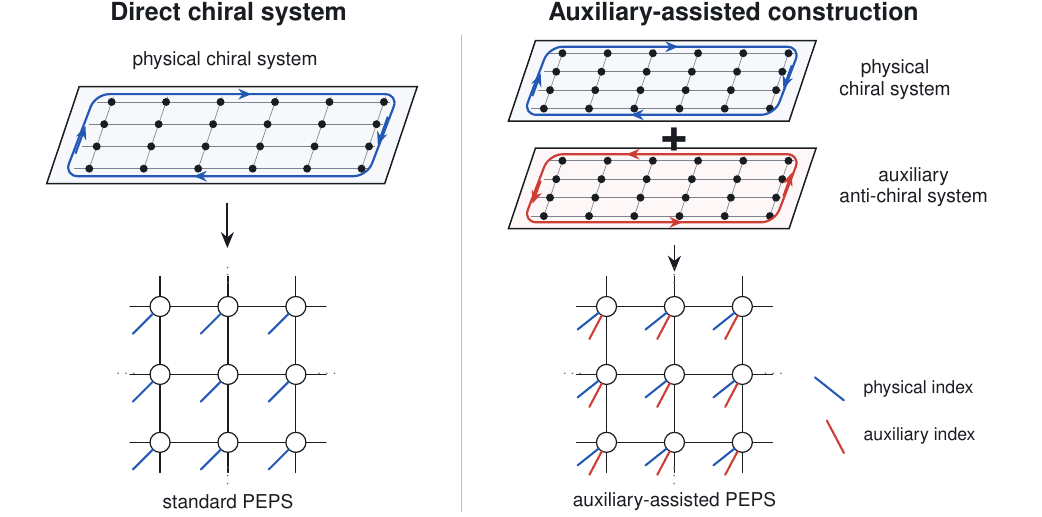}
\caption{\emph{\textbf{Sketch diagrams for direct and auxiliary-assisted PEPS representations.}} In the auxiliary-assisted construction, the physical chiral system and its anti-chiral auxiliary partner have separate physical indices but share the same virtual bonds. The emergent finite-$D$ entanglement between the two subsystems is crucial for bypassing the chiral PEPS obstruction.}
\label{fig:sketch}
\end{localgraphicspath}
\end{figure}

Here we take a qualitatively different route. Rather than seeking a finite-$D$ PEPS within the Hilbert space of the original chiral system, we consider the original chiral system together with an independent and completely decoupled time-reversed partner, see Fig.~\ref{fig:sketch}. Although the combined system has no net chirality, this pairing alone does not remove the obstruction: if the finite-$D$ PEPS remained factorized between the two sectors, it would still reduce to two separately obstructed chiral representations (see proof in Supplemental Material~\cite{ref_appendix}).
One might therefore expect that an explicit intersector coupling term  must be added to the Hamiltonian to regularize the representation. 
Remarkably, we find that this is unnecessary. Finite-$D$ variational optimization itself generates an \emph{emergent} residual entanglement between the two sectors, which removes the spurious long-range correlations of the isolated chiral PEPS, despite the absence of any explicit intersector coupling in the Hamiltonian.

%We demonstrate this approach for both a free-fermion Chern insulator and an interacting spin-$1/2$ $J_1-J_2-J_\chi$ chiral spin liquid. In both cases, the auxiliary-assisted PEPS exhibits clean short-range correlations, while a direct chiral representation retains the characteristic long-distance tail. For the free-fermion model, a same-chirality bilayer provides a control showing that the improvement originates from cancellation of chirality rather than simply from enlarging the local Hilbert space. Finally, although the enlarged state is nonchiral, its chiral edge information remains accessible. Approximately layer-resolved momentum projection of the entanglement spectrum recovers the expected low-lying universal conformal-tower structure. These results establish an auxiliary-assisted finite-$D$ PEPS framework that bypasses the obstruction to a direct short-ranged PEPS representation of the chiral pure state, while retaining access to its gapped bulk physics and universal chiral information.

We demonstrate this mechanism for both a free-fermion Chern insulator and an interacting spin-$1/2$ chiral spin liquid. In both cases the emergent physical-auxiliary entanglement eliminates the long-distance correlation artifact, while local density matrices increasingly factorize between the two sectors with increasing $D$. Although the enlarged state is nonchiral, its chiral edge information remains accessible. Layer-resolved momentum projection of the entanglement spectrum recovers the expected low-lying universal conformal-tower structure. 
%These results establish an auxiliary-assisted PEPS framework that bypasses the obstruction to a direct short-ranged PEPS representation of the chiral states, while retaining access to their gapped bulk physics and universal chiral information.
Together, these results uncover a previously unrecognized variational mechanism by which auxiliary entanglement bypasses the obstruction to a direct short-ranged PEPS representation of chiral states, while preserving their gapped bulk physics and universal chiral information.

{\bf \emph{Framework: An auxiliary-assisted wavefunction for chiral physics.}} Our construction embeds the physical chiral system into an enlarged non-chiral Hilbert space by introducing a decoupled auxiliary anti-chiral model. Specifically, we consider
\begin{equation}
    H_{\rm total} = H_{\mathrm{phy}} \otimes I_{\mathrm{aux}} + I_{\mathrm{phy}} \otimes H_{\mathrm{aux}},
\end{equation}
including the original \emph{physical} chiral system $H_{\mathrm{phy}}$ and its anti-chiral time-reversal partner $H_{\mathrm{aux}}=\Theta H_{phy}\Theta^{-1}$ as an \emph{auxiliary} system. Here $I$ denotes the identity operator. 
%The local Hilbert space and its dimension are enlarged from $({\mathcal H_{phy}}, d_{\rm phy})$ to 
%\begin{equation}
%    \mathcal H_{\mathrm{tot}} = \mathcal H_{\mathrm{phy}} \otimes \mathcal H_{\mathrm{aux}}, \qquad d_{\rm total} = d_{\rm aux}d_{\rm phy}=d_{\rm phy}^2.
%\end{equation}
The local Hilbert space dimension is extended from $d$ to $d_{\rm total}=d^2$.
%The exact ground state therefore factorizes as the tensor product of the physical and auxiliary parts. Nevertheless, the enlarged system has no net chirality and can be represented efficiently by a finite-$D$ PEPS. 
The exact ground state factorizes, but a factorized finite-$D$ PEPS would still consist of two separately obstructed chiral representations~\cite{ref_appendix}. The variational PEPS is therefore allowed, but not forced, to develop interlayer entanglement.

We optimize an infinite PEPS~\cite{ctmrg1,ctmrg2,orus2009simulation,liao2019differentiable} for $H_\mathrm{total}$ with a local tensor $A^{s_{\rm aux}s_{\rm phy}}_{LDRU}$, with $s_{\rm phy}$ and $s_{\rm aux}$ retaining the tensor-product structure, while $L,D,R,U$ denote common virtual indices of a finite bond dimension $D$. This distinction is central to the construction. 
Although the microscopic layers are decoupled, they share the same finite-dimensional virtual entanglement space.
%The variational tensor is therefore not constrained to factorize into independent physical and auxiliary PEPS tensors.

%Consequently, at finite $D$ the optimized state generally satisfies
%\begin{equation}
 %   \vert \psi_{\rm tot} \rangle \neq \vert \psi_{\rm phy} \rangle \otimes \vert \psi_{\rm aux} \rangle
%\end{equation}
Consequently, at finite $D$ the optimized state generally satisfies $ \vert \psi_{\rm tot} \rangle \neq \vert \psi_{\rm phy} \rangle \otimes \vert \psi_{\rm aux} \rangle$
and contains weak residual correlations between the two layers. These correlations provide the additional variational freedom that distinguishes the enlarged non-chiral representation from a direct finite-$D$ representation of an isolated chiral state. In this sense, the finite-entanglement approximation pays the price of a small violation of layer factorization rather than producing the artificial long-range correlations characteristic of a finite-$D$ chiral PEPS.

A useful phenomenological description of this finite-$D$ state is 
\begin{equation}
    H_{\text{eff}} = H_{\rm total} + \sum_i \lambda_i \, O^i_\mathrm{phy} \otimes O^i_\mathrm{aux} 
    \label{eq:Heff}
\end{equation}
where $\lambda_i$ are effective local couplings associated with the residual physical-auxiliary entanglement at finite $D$. 
%This is similar to the entanglement induced relevant deformations of a quantum critical point~\cite{finite_entanglement_CFT}. 
For a gapped system, sufficiently weak local couplings generally can not induce a phase transition. We therefore expect low-energy properties in the physical sector are close to that in the pure chiral physical system especially when the gap is large, while the residual couplings only affect short-distance correlations and decrease systematically when enlarging entanglement space.

Physical quantities are obtained by tracing out the auxiliary layer $\rho_{\rm phy} = \Tr_{\rm aux}{\vert \psi_{\rm tot}} \rangle \langle \psi_{\rm tot} \vert$. Local observables, correlation functions, and finite-region reduced density matrices of PEPS are evaluated directly in the thermodynamic limit by corner-transfer-matrix renormalization group (CTMRG)~\cite{ctmrg1,ctmrg2,orus2009simulation} with environment bond dimension $\chi$. The tensors are optimized variationally, with gradients computed by automatic differentiation~\cite{liao2019differentiable}.

%In the following we are going to show that, indeed as expected the chiral correlation function shows correct clean gapped behaviors, especially in the long range $r \gg \xi$ with $\xi$ being the correlation length. More importantly, the entanglement induced residual couplings only affect short-distance correlations and do not change low-energy physical and topological properties of the original physical system. 

{\bf \emph{Gapped correlations.}} 
%We first demonstrate that the auxiliary construction removes the artificial long-range correlations that appear in a direct finite-$D$ PEPS representation of chiral states. 
We first demonstrate the qualitative consequence of the emergent entanglement: the artificial long-range correlations of a direct finite-$D$ chiral PEPS are removed.
We consider two representative examples: a free-fermion Chern insulator and an interacting chiral spin liquid. For the fermionic case, we consider the triangular-lattice $\pi/2$-flux Chern insulator of Refs.~\cite{hofstadter1976energy,avron2014study,kuhlenkamp2024chiral}, denoted by $H_{\rm CI}$, whose two bands carry $C=\pm1$ (see SM), and use the standard fermionic PEPS formalism~\cite{barthel2009contraction,kraus2010fermionic,gu2563grassmann,pivzorn2010fermionic,corboz2010simulation,gu2013efficient}. In the double-layer setup, we take $H_{\mathrm{phy}}=H_{\rm CI}$ and $H_{\mathrm{aux}}=-H_{\mathrm{CI}}$, so that the auxiliary layer carries the opposite chirality.

To show the effect of our construction, we compare it with two different setups for the fermionic model. The first one is the direct original chiral fermion model $H_\mathrm{CI}$, for which it was known that PEPS fail to describe the correct correlation functions. The other one is an identical second copy of the physical model $H_\mathrm{CI}$, dubbed ``chiral+chiral''. One may assign opposite spin labels for the two layers, hence the model can be viewed as a $\rm{SU}(2)$-symmetric spinful model in which each component carries the same chirality. %It is typically used in $\rm{SU}(2)$-symmetric Hubbard-type problems. 
For the bi-layered models, the total onsite Hilbert space is enlarged, therefore we compare them with the singe layer chiral model (spinless) using the square root $\sqrt{\tilde D}$ of the virtual bond dimension $\tilde D$. 

\begin{figure}[hbt!]
\begin{localgraphicspath}{{figs/}}
\includegraphics[width=1\columnwidth]{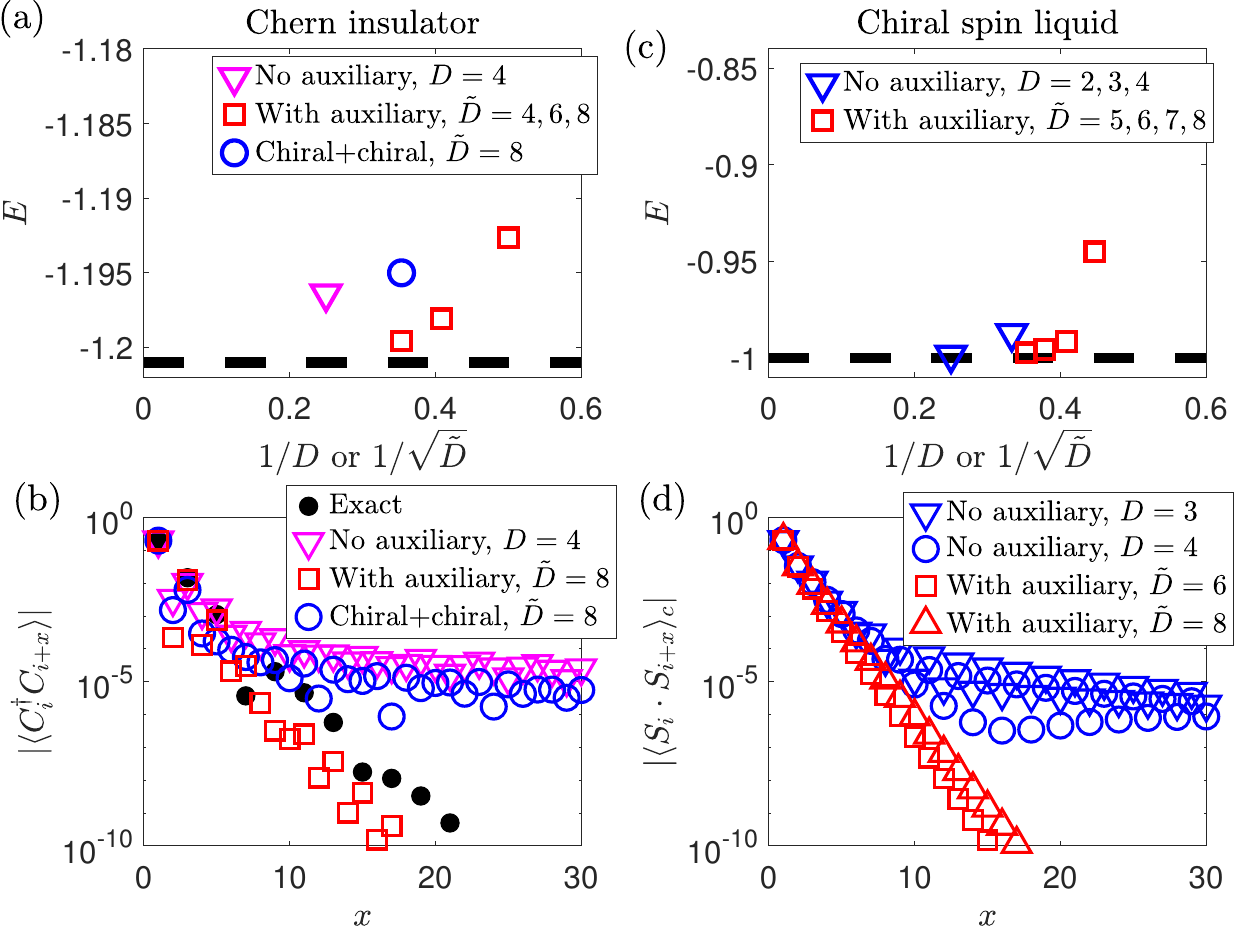}
\caption{\emph{\textbf{Variational energies and correlation functions with and without antichiral auxiliary systems.}}  Variational energy (a) (c) and correlation functions (b) (d) of the optimized infinite PEPS states for both models are presented. The exact free-fermion energy density and the accurate reference energy for the spin model are also shown. In (a)-(b), ``chiral+chiral'' model refers to the bi-layer system where the Hamiltonian in the auxiliary layer is identical to that of the physical layer. For better comparison with the original single layer system, we used the square root of the bond dimension $\sqrt{\tilde{D}}$ for the bilayered calculation in the plot.}
\label{fig:energy_correlation}
\end{localgraphicspath}
\end{figure}

The three setups already show different variational behavior in energy. As shown in Fig.~\ref{fig:energy_correlation}(a), the auxiliary-assisted setup converges more smoothly when plotted against $1/\sqrt{\tilde D}$, suggesting a less obstructed variational representation. The decisive test is provided by the long-distance correlation functions. The direct spinless case and the ``chiral+chiral'' cases reproduce the short-distance behavior but develops a long-distance tail artifact [Fig.~\ref{fig:energy_correlation}(b)]. In contrast, the auxiliary-assisted setup exhibits a clean exponential decay over the accessible distance range. %Thus, the correlation artifact is tied to the uncompensated chirality and disappears when the two spin components carry opposite chiralities.
Thus, the long-distance correlation artifact persists in the single-chiral and chiral $+$ chiral cases, but is eliminated in the presence of an opposite-chirality auxiliary.

We then turn to the interacting chiral spin liquid.
The spin model is a $J_1-J_2-J_{\chi}$ model defined by the $\rm{SU}(2)$ spins on the square lattice~\cite{nielsen2013local,poilblanc2017}
\begin{equation}\label{eq:hamiltonian}
{\cal H_{\rm{spin}}} = J_1 \sum_{\langle i,j \rangle} {\bf S}_i \cdot {\bf S}_j + J_2 \sum_{\langle \langle i,k \rangle \rangle} {\bf S}_i \cdot {\bf S}_k + J_{\chi} \sum_{\triangle_{ijk}} \left({\bf S}_i \times {\bf S}_j \right) \cdot {\bf S}_k.
\end{equation}
We choose the parameters as $ J_1=2\cos(0.06\pi)\cos(0.14\pi)$, $ J_2=2\cos(0.06\pi)\sin(0.14\pi)$, $J_{\chi}=4\sin(0.06\pi)$ with the ground state energy being $E\approx  -1$~\cite{poilblanc2017,hasik2022}. The auxiliary system is defined as the time-reversal of $H_{\rm spin}$ with a reversed sign of $J_{\chi}$.

For the CSL simulation, both approaches reach comparable variational energies, but the auxiliary construction approaches the reference energy more rapidly when plotted against $1/\sqrt{\tilde D}$, compared with the direct single-layer scaling in $1/D$ [Fig.~\ref{fig:energy_correlation}(c)]. The decisive difference again appears in the spin correlations: the direct chiral PEPS develops the characteristic gossamer tail, whereas the auxiliary construction shows a purely short-ranged decay with no visible power-law-like tail [Fig.~\ref{fig:energy_correlation}(d)]. Thus, the suppression of the spurious long-distance tail extends beyond free-fermion Chern bands to an interacting chiral topological order.

\begin{figure}[hbt!]
\begin{localgraphicspath}{{figs/}}
\includegraphics[width=1\columnwidth]{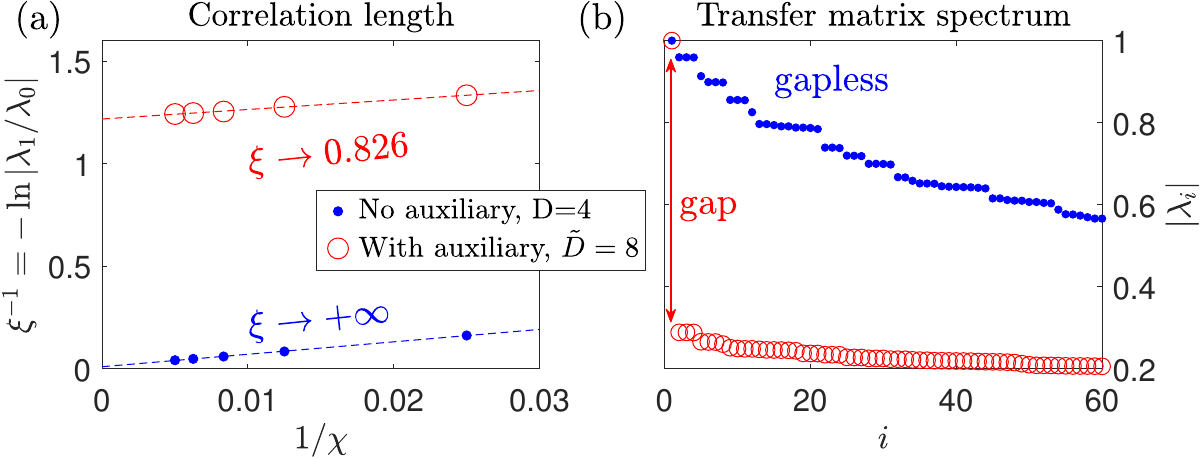}
\caption{\emph{\textbf{Correlation length of the variational states in spin model.}}  (a) Extrapolation of correlation length versus $1/\chi$. (b) Spectrum of transfer matrix for computing correlation functions with environment dimension $\chi=200$. }
\label{fig:correlation_length}
\end{localgraphicspath}
\end{figure}

{\bf \emph{Finite correlation length.}}~The correlation functions above demonstrate that the long-range tail is removed. We now examine the correlation length directly using the transfer matrix of the auxiliary-assisted CSL state. The correlation length $\xi$ of a PEPS can be conveniently obtained as
\begin{align}
\xi = -1/\log |\lambda_1/\lambda_0|,
\end{align}
where $\lambda_0$ and $\lambda_1$ are the leading and sub-leading eigenvalues of the transfer matrix. Figure~\ref{fig:correlation_length}(a) shows the extrapolation of $\xi^{-1}$ with the environment dimension $\chi$. For the single chiral PEPS, $\xi^{-1}$ extrapolates toward zero, corresponding to an artificial diverging correlation length. In contrast, the auxiliary-extended state extrapolates to a short correlation length $\xi\sim 0.826$, demonstrating a genuinely short-ranged variational state. Consistently, the transfer-matrix spectrum in Fig.~\ref{fig:correlation_length}(b) is nearly gapless for the single chiral PEPS but clearly gapped for the auxiliary-extended state, confirming that the exponential real-space decay reflects a finite transfer-matrix correlation length.

\begin{figure}[hbt!]
\begin{localgraphicspath}{{figs/}}
\includegraphics[width=1\columnwidth]{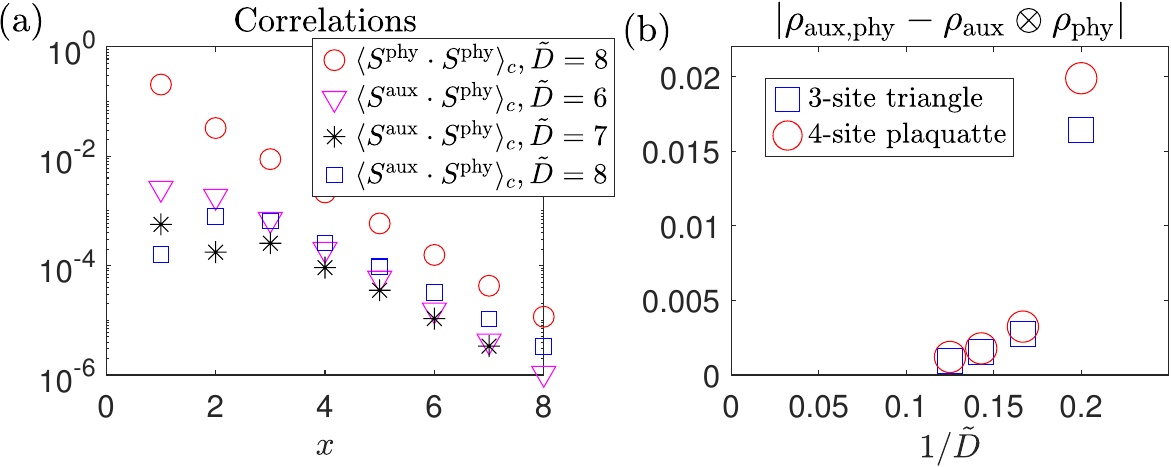}
\caption{\emph{\textbf{Residual physical-auxiliary correlations and factorization error}}. (a) Comparison of spin-spin correlation function between physical-physical and auxiliary-physical. (b) Frobenius norm of the reduced-density-matrix difference measuring the deviation from physical-auxiliary factorization. }
\label{fig:CSL_decoupling}
\end{localgraphicspath}
\end{figure}

{\bf \emph{Residual physical-auxiliary entanglement.}}~At finite bond dimension, the optimized PEPS generally contains residual weak physical-auxiliary correlations. We first examine spin correlations between the two layers. The nonzero physical-auxiliary correlations shown in Fig.~\ref{fig:CSL_decoupling} (a) are consistent with the effective weak-coupling picture of Eq.~\ref{eq:Heff}. These correlations  are strongly suppressed compared with the physical-physical spin correlations and decay exponentially with distance for all bond dimensions studied. Thus the auxiliary system does not generate long-range physical correlations in the physical chiral system.

A more systematic measure of decoupling is obtained from finite-cluster reduced density matrices. For a finite region $\mathcal R$ on the infinite lattice, we define the factorization error
\begin{equation}
\Delta_{\mathcal R} =
\left\|
\rho_{\mathcal R}^{\rm aux, phy}-\rho_{\mathcal R}^{\rm aux}\otimes \rho_{\mathcal R}^{\rm phy}
\right\|_F ,
\end{equation}
where $\|\cdot\|_F$ denotes the Frobenius norm. This is stronger than any single correlation function, since it probes the reduced density matrix on $\mathcal R$ and tests whether it approximately factorizes between the physical and auxiliary subsystems. Figure.~\ref{fig:CSL_decoupling} (b) demonstrates the factorization error decreases rapidly with increasing bond dimension. 
%Together with the weak short ranged inter-layer correlations, our studies show that the residual physical-auxiliary coupling in the entanglement space only affects short-distance quantities and can be systematically suppressed when enlarging $D$.
Together with the weak short-ranged interlayer correlations, our results show that the residual physical-auxiliary entanglement manifests only through short-distance correlations and is systematically suppressed with increasing $D$.

These results confirm the weak-coupling picture underlying our construction. The auxiliary layer provides the additional entanglement structure needed to remove the chiral PEPS obstruction, while introducing only weak and short-ranged correlations with the physical system. As we show next, the residual entanglement is sufficiently weak that the chiral topological information of the physical system remains resolvable.

{\bf \emph{Resolving chiral edge physics in the enlarged state.}} Two-dimensional chiral gapped phases exhibit characteristic bulk-boundary correspondence. Their topological data are reflected in an anomalous gapless (1+1)-dimensional boundary theory~\cite{wen_chiral_1990,wen_chiral_1991,elitzur1989remarks,li2008entanglement,qi_ent_top_2012}.
According to Li-Haldane conjecture~\cite{li2008entanglement,qi_ent_top_2012}, their universal chiral edge structure is encoded in the low-lying entanglement spectrum. 
We thus place the state on an infinite cylinder and split it into two half-infinite cylinders, producing a single entanglement boundary around the circumference~\cite{cirac2011entanglement,poilblanc20162}. 
For the CSL considered here, the chiral $\rm{SU}(2)$ edge theory~\cite{di1997conformal,blumenhagen2012introduction} has two topological ground-state sectors on an infinite cylinder, labeled by the vacuum $I$ and semion $s$ fluxes, whose entanglement spectra realize the corresponding $I$ and $s$ conformal towers, respectively.
The low-energy entanglement Hamiltonian and momentum are governed by the same chiral conformal generator, $H_E\sim L_0$ and $P_y\sim L_0$~\footnote{Nonuniversal velocities, normalization factors, and constant shifts are suppressed here}, so that each primary sector generates a unidirectional branch organized into a conformal tower~\cite{ref_appendix}. 
In our enlarged construction, the physical chiral system is combined with an anti-chiral partner, and the enlarged entanglement boundary is described instead by a non-chiral $\rm{SU}(2)_1\times\overline{\rm{SU}(2)}_1$ theory, with $H_E\sim L_0+\bar L_0$ and $P_y^{\rm tot}\sim L_0-\bar L_0$. 

The important point, however, is that this chirality cancellation does not erase the individual chiral components. The physical and auxiliary degrees of freedom remain explicitly distinguished in the microscopic Hilbert space, $\mathcal H_{\rm phy} \otimes \mathcal H_{\rm aux}$. As we analyzed above, the weak mixing only affects the gapped bulk in a short range and controlled way, which is naturally expected to transform to irrelevant deformation to the boundary CFT which do not modify the universal low-energy structure. One may therefore approximately translate the two layers independently. The corresponding layer momenta resolve the two conformal levels separately, $k_y^{\rm phy} \sim h+N, \, k_y^{\rm aux} \sim -(\bar h+\bar N)$
while their simultaneous translation gives $k_y^{\rm phy} + k_y^{\rm aux} \sim
h+N-\bar h-\bar N$~\cite{ref_appendix}.

\begin{figure}[hbt!]
\begin{localgraphicspath}{{figs/}}
\includegraphics[width=1\columnwidth]{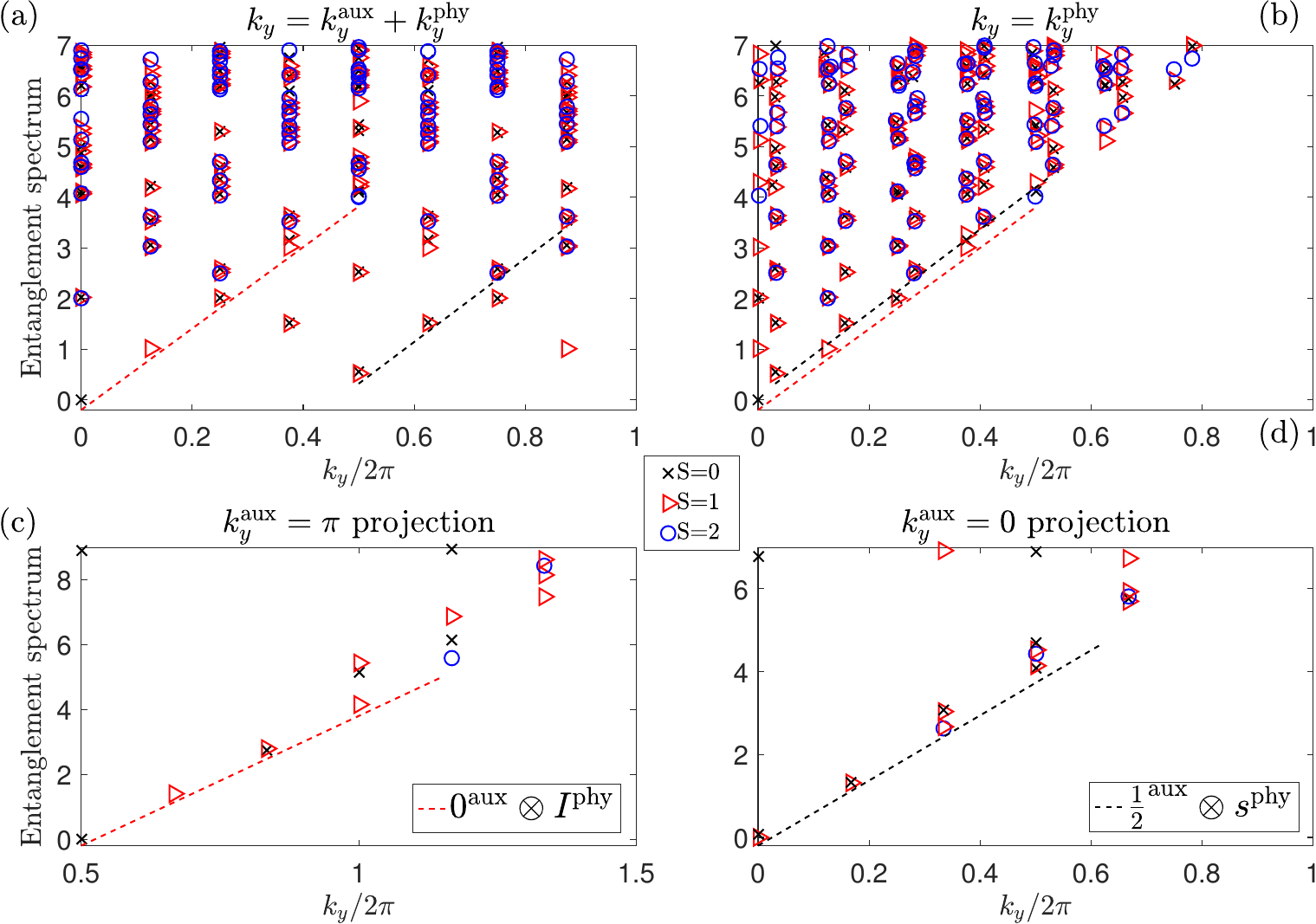}
\caption{\emph{\textbf{Bipartite entanglement spectrum of the spin state on finite-width cylinders.}} Variational $\tilde{D}=6$ state (with a virtual space $V=0\oplus 0\oplus \frac12\oplus \frac12$) with $\rm{SU}(2)$ symmetry. (a)-(b) Entanglement spectrum of the enlarged physical-plus-auxiliary wavefunction, resolved by the total momentum in (a) and by the physical momentum in (b) on a width-8 cylinder. CTMRG environment with $\chi=80$ is used to construct entanglement Hamiltonian. (c)-(d) Projected spectra in fixed auxiliary-momentum sectors on a width-6 cylinder. Exact contraction is used to construct entanglement Hamiltonian. }
\label{fig:ES}
\end{localgraphicspath}
\end{figure}

Resolving the Schmidt states by the total momentum ($k_y^{\rm tot}$) gives the ordinary non-chiral spectrum of the enlarged state, as shown in Fig.~\ref{fig:ES} (a). Instead resolving with respect to the physical-layer momentum ($k_y^{\rm phy}$) reorganizes the same low-energy states according to the chiral conformal level $h+N$, in Fig.~\ref{fig:ES} (b). For a state belonging to the conformal sectors ($i,j$), $\epsilon_E \sim
h_i+N+\bar h_j+\bar N$, whereas $k_y^{\rm phy} \sim h_i+N$. The auxiliary level therefore enters only as an additive offset, $\epsilon_E
\simeq k_y^{\rm phy} + \epsilon_{\rm aux}$. Different auxiliary levels consequently generate different approximately parallel chiral branches. The multiple dispersing branches in Fig.~\ref{fig:ES} (b) should thus be viewed not as independent physical edge theories, but as copies of the physical chiral tower dressed by different auxiliary entanglement states. The topological-sector structure further constrains these branches. 

Since the $I$ and $s$ towers contain integer and half-integer $\rm{SU}(2)$ multiplets, respectively, the universal low-energy spectrum in the integer-spin Schmidt sector [Fig.~\ref{fig:ES}] is restricted to the diagonal combinations $\mathcal{H}_{\mathrm{integer}} = (\bar I^{\rm aux}\otimes I^{\rm phy}) \oplus (\bar s^{\rm aux}\otimes s^{\rm phy})$.
The dashed lines indicate the low-lying branches that are further isolated by the auxiliary-momentum projections in Figs.~\ref{fig:ES} (c)-(d), which is achieved by projecting onto fixed auxiliary-momentum sectors. The $k_y^{\rm aux}=\pi$ and $k_y^{\rm aux}=0$ projections select the lowest auxiliary level in the $\bar I^{\rm aux}\otimes I^{\rm phy}$ and $\bar s^{\rm aux}\otimes s^{\rm phy}$ towers, yielding the single $0^{\rm aux}\otimes I^{\rm phy}$ and $\frac{1}{2}^{\rm aux}\otimes s^{\rm phy}$ branches, respectively. Now the auxiliary conformal level is fixed and the entanglement spectrum reduces to the chiral branch $\epsilon_E \sim k_y^{\mathrm{phy}}$. The low-lying $\rm{SU}(2)$ multiplet counting is also consistent with the CFT~\cite{ref_appendix}. 
These results show that the emergent auxiliary entanglement bypasses the finite-$D$ chiral obstruction without erasing the chiral topological information: the latter remains encoded in the enlarged entanglement spectrum and can be extracted by subsystem-momentum resolution.

{\bf \emph{Conclusions.}} 
In this work, we introduce an auxiliary anti-chiral system-assisted framework and uncover an emergent variational mechanism that bypasses the long-standing obstruction to short-ranged PEPS representations of gapped chiral states. Although the physical and auxiliary Hamiltonians are completely decoupled, variational optimization generates weak residual interlayer entanglement that regularizes the PEPS representation, removes the spurious gapless long-range correlations, and restores the genuinely short-ranged character of the target phase. This residual entanglement is systematically suppressed with increasing $D$, while the universal chiral information remains accessible through the layer-resolved entanglement spectrum.

%The mechanism is closely analogous to finite entanglement scaling of quantum critical point~\cite{finite_entanglement_CFT}. The PEPS wave function is not a direct tensor product of the physical and auxiliary system. Weak residual auxiliary-physical couplings are generated in the entanglement space, which regularize the PEPS representation. These weak couplings only affect short range or high energy physical quantities, leaving the long-range physics and more importantly universal topological feature of the chiral physical system unchanged. Our framework of using an auxiliary system provides a practical working method to study chiral gapped physics. 

This mechanism is reminiscent of finite-entanglement scaling at quantum critical points~\cite{finite_entanglement_CFT}: the finite entanglement of the variational state generates a weak deformation that regularizes the representation and vanishes as the variational space is enlarged. Here this deformation takes the form of short-ranged physical-auxiliary entanglement, while the universal chiral structure remains intact.

There are quite a few interesting points we leave for future studies. Our study provides a window to study how the chiral and anti-chiral part in a non-chiral system becomes inseparable. Starting from local operator contents in a chiral gapped theory and its anti-chiral one, a systematic classification of possible coupling can be made and analyzed. It is also interesting to explore how a small irrelevant coupling makes a true gapped PEPS wave function. The auxiliary-assisted construction also suggests a useful connection with purified mixed-state representations. Due to the finite entanglement effect, tracing out the auxiliary layer produces a weakly mixed physical state, which may be viewed heuristically as analogous to a low-temperature deformation of the target ground-state density matrix. The analogy indicates that sufficiently weak physical–auxiliary entanglement should leave low-energy local physics essentially unchanged, similarly to thermal corrections in a gapped system when $\beta\,\Delta \gg 1$. Developing a genuine finite-temperature or purified-density-matrix formulation of chiral phases is an interesting direction for future work.

{\bf \emph{Acknowledgements.}} We thank Xiao-Han Yang, Yi-Ming Liu, Wen-Tao Xu, Sasank Budaraju and Chao Xu for helpful discussions. S.N. is supported by the National Natural Science Foundation of China (Grant No. 12604217). R.Z.H. is supported by the National Natural Science Foundation of China (Grant No. 12504183).
\bibliography{ref}

\newpage
\clearpage
%\appendix
%\widetext
\onecolumngrid

In this Supplemental Material (SM), we present a proof that exact physical-auxiliary factorization does not remove the chiral PEPS obstruction, details of the microscopic Hamiltonian, and technical details of the entanglement-spectrum extraction and bulk-entanglement boundary correspondence.

\section{Why a factorized bilayer PEPS remains obstructed}

Here we show that simply combining a chiral system with a decoupled anti-chiral auxiliary does not remove the chiral PEPS obstruction if the variational state remains exactly factorized. The argument can be summarized in four steps.

\begin{enumerate}

\item \textit{Assume a short-ranged factorized finite-$D$ PEPS.}

Consider a finite system of arbitrary size and suppose that the enlarged bilayer admits a short-ranged finite-$D$ PEPS representation that factorizes exactly between the physical and auxiliary sectors,
\begin{equation}
    |\Psi_{\rm fact}\rangle
    =
    |\Psi_{\rm phy}\rangle
    \otimes
    |\Psi_{\rm aux}\rangle .
\end{equation}

\item \textit{Exact factorization preserves all physical correlation functions.}

For any operators $O_i$ and $O_j$ acting only on the physical layer, exact factorization gives
\begin{equation}
    \langle O_i O_j\rangle_{\Psi_{\rm fact}}
    =
    \langle O_i O_j\rangle_{\Psi_{\rm phy}},
\end{equation}
and similarly
\begin{equation}
    \langle O_i O_j\rangle_{c,\Psi_{\rm fact}}
    =
    \langle O_i O_j\rangle_{c,\Psi_{\rm phy}} .
\end{equation}
Therefore, if the factorized bilayer PEPS has exponentially decaying correlations, the isolated physical state $|\Psi_{\rm phy}\rangle$ must have the same short-ranged correlation behavior.

\item \textit{A local projection onto the auxiliary sector preserves finite bond dimension.}

Choose an auxiliary product configuration
\begin{equation}
    |s_{\rm aux}\rangle
    =
    \bigotimes_i |s_i\rangle
\end{equation}
with
\begin{equation}
    \langle s_{\rm aux}|\Psi_{\rm aux}\rangle \neq 0 .
\end{equation}
Such a configuration always exists in a finite system since $|\Psi_{\rm aux}\rangle$ has a nonzero amplitude in at least one product-basis configuration. Projecting the auxiliary physical indices gives
\begin{equation}
    \langle s_{\rm aux}|\Psi_{\rm fact}\rangle
    =
    \langle s_{\rm aux}|\Psi_{\rm aux}\rangle
    |\Psi_{\rm phy}\rangle .
\end{equation}
The projection $\langle s_{\rm aux}|$ is a product of local rank-one maps and therefore does not increase the virtual bond dimension. Hence $|\Psi_{\rm phy}\rangle$ itself admits a finite-$D$ PEPS representation.

\item \textit{Contradiction with the chiral PEPS obstruction.}

Combining the previous two results, a short-ranged factorized finite-$D$ PEPS for the enlarged bilayer would imply a short-ranged finite-$D$ PEPS representation of the isolated chiral state $|\Psi_{\rm phy}\rangle$. This contradicts the chiral PEPS obstruction of the original physical sector.

\end{enumerate}

Therefore, cancellation of the total chirality in a decoupled bilayer is not sufficient to remove the obstruction. A short-ranged finite-$D$ representation of the enlarged system must depart from exact physical-auxiliary factorization. This establishes that interlayer entanglement is a necessary ingredient of the auxiliary-assisted representation; our variational results show that precisely such entanglement emerges spontaneously at finite $D$.

\section{Hamiltonian of free-fermion Chern insulator}\label{app_0}

The fermionic model is a nearest neighbour tight-binding model on the triangular lattice pierced by $\pi/2$ flux for each triangle 
\begin{align}
H_{\rm{CI}}&=\sum_{\langle i,j\rangle} (t_{ij}c_{i}^{\dagger} c_{j}+\rm{h.c.}).
\label{eq:H} 
\end{align}
We adopt the gauge such that the complex nearest neighbour hoppings along the three primitive translation vectors $\boldsymbol{e}_1=(1,0)$, $\boldsymbol{e}_2=(-1/2,\sqrt{3}/2)$, $\boldsymbol{e}_3=\boldsymbol{e}_1+\boldsymbol{e}_2$ take the form $t_{\boldsymbol{r},\boldsymbol{r}+\boldsymbol{e}_1}=it$, $t_{\boldsymbol{r},\boldsymbol{r}+\boldsymbol{e}_2}=(-1)^{x}t$, $t_{\boldsymbol{r},\boldsymbol{r}+\boldsymbol{e}_3}=(-1)^{x-1}t$ ~\cite{kuhlenkamp2024chiral}. 
In our double-layer setup, the physical subsystem is defined as $H_{\mathrm{phy}}=H_{\rm{CI}}$, and the auxiliary system is chosen as $H_{\mathrm{aux}}=-H_{\mathrm{CI}}$ which is the anti-chiral variant of $H_\mathrm{phy}$ up to a local gauge transformation. The total decoupled Hamiltonian mimics a quantum spin Hall system.

\section{Extract Entanglement Spectrum on a finite Cylinder}\label{app_1}
Consider a system on an infinite long cylinder with a finite width $W$ and its half-infinite subsystem $A$. 
For PEPS, the corresponding entanglement Hamiltonian is naturally related to the boundary theory of the virtual degrees of freedom~\cite{cirac2011entanglement}.
The subsystem A and the rest $\overline{A}$ are separated by a single cut. The bipartite entanglement Hamiltonian $H_E$ for A is defined from the reduced density matrix
\begin{equation}
\rho_A = \frac{e^{-H_E}}{\mathrm{Tr}\, \left ( e^{-H_E} \right)}    
\end{equation}
in which the spectrum of $H_E$ can be understood as the energy spectrum of the entanglement boundary theory. In our work, we use PEPS wave function $\vert \psi \rangle$ to approximate the ground state and then we put the wave function on the infinite long cylinder. Here we are using pure state to study the entanglement, hence a more convenient and equivalent way to extract spectrum of $H_E$ is from the singular value decomposition
\begin{equation}
\vert \,\psi_{A\,\overline{A}} \rangle = \sum_{j} \lambda_j \, |u_{A,j}\rangle \otimes \vert v_{\overline{A},\, i}\rangle,    
\end{equation}
\label{eq:H}
where $|u_{A,j}\rangle$ and $|v_{\overline{A},\,j}\rangle$ form orthogonal basis for the subsystem $A$ and $\overline{A}$, respectively. $\lambda_j$ gives spectrum $\Delta_i$ of $H_E$ through $\Delta_i \sim -\log(\lambda_i^2)$.

Along the finite width $y$ direction, there is translation symmetry. One can introduce translation operators $T_{y}$ along $y$ direction, which commute with the bipartite reduced density matrix. Thus by diagonalizing the translation operator and the reduced density matrix simultaneously, one can label each Schmidt basis state with a momentum
\begin{equation}
T_{y} \, \vert u_{A,j}\rangle = e^{i\,k_{j} a}|u_{A,j}\rangle,
\end{equation}
where $k_{j}$ is the momentum.

\begin{figure}[hbt!]
\begin{localgraphicspath}{{figs/}}
\includegraphics[width=0.8\columnwidth]{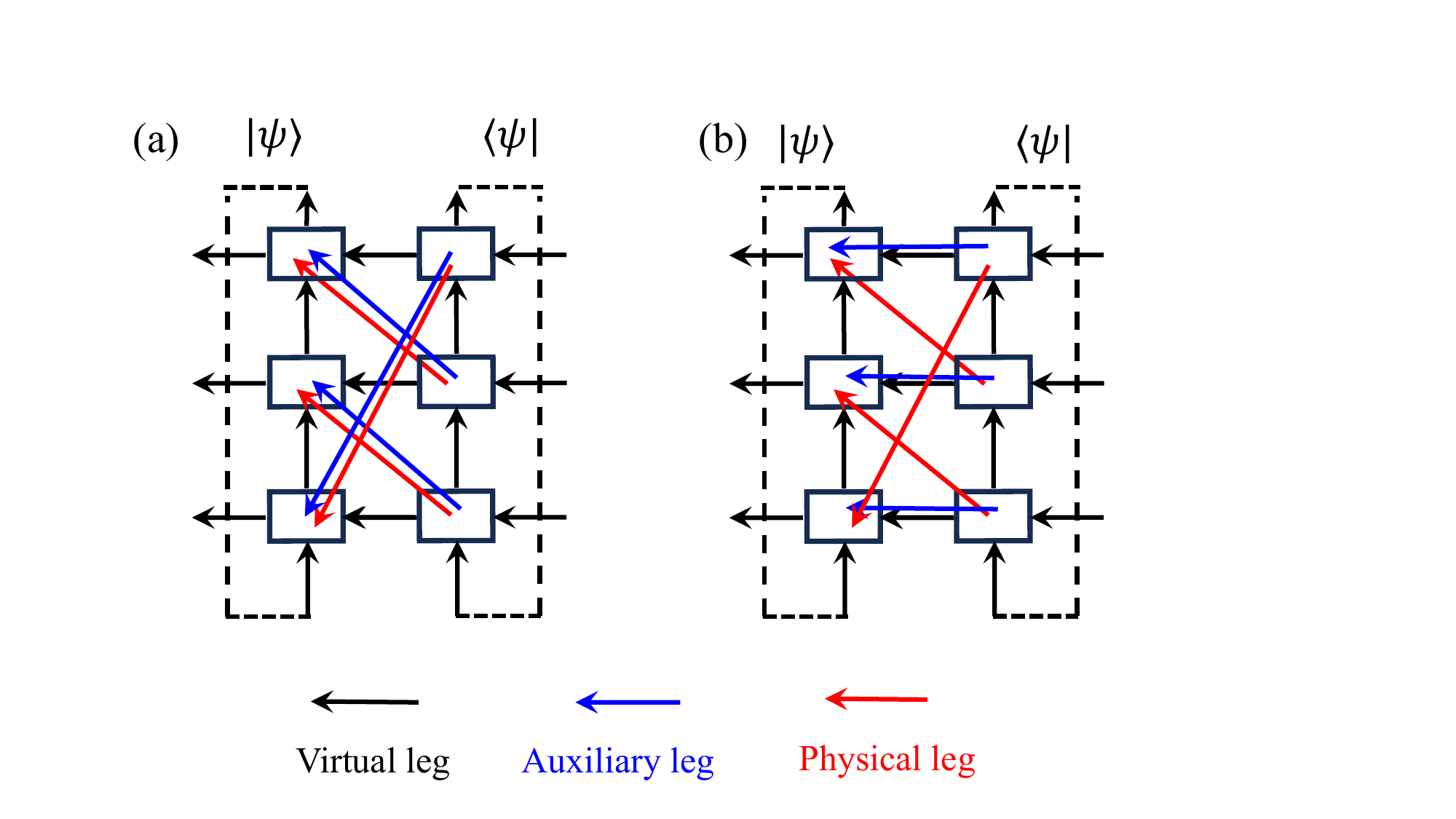}
\caption{Sketch diagram for translation operator actions used in the calculation of entanglement spectrum on a $W=3$ cylinder. (a) Full ${T}_y={T}_{y}^{\rm aux} \, {T}_{y}^{\rm phy}$ translation for both physical and auxiliary systems. (b) Partial ${T}_y={T}_{y}^{\rm phy}$ translation for only the physical system. }
\label{fig:tensor_diagram}
\end{localgraphicspath}
\end{figure}

Our bi-layered model contains a physical and auxiliary system. In the variational optimization we obtain the PEPS wave function for the whole system, in which we can not write the wave function as a direct tensor product between a physical and auxiliary parts -- it is actually the weak residual coupling between these two systems in the entanglement degrees of freedom removes the spurious long-range chiral correlation as we analyzed in the main text. Nevertheless, we can approximately use the translation invariance in each layers and define two independent momentum for the two layers respectively. In the decoupled limit of our double layer framework, the whole wavefunction takes the form of
\begin{equation}
    \vert \psi_{\mathrm{tot}}\rangle = \vert \psi_{\mathrm{phy}}\rangle \otimes \vert \psi_{\mathrm{aux}}\rangle.
\end{equation}
and 
\begin{equation}
|\psi_{\rm{tot}}\rangle = \sum_{m,n} \lambda_m \lambda_n \, \left(|u_{A,m}\rangle_{\rm phy} \otimes |\bar{u}_{A,n}\rangle_{\rm aux} \right)\otimes \left(|v_{\overline{A},m}\rangle_{\rm phy} \otimes |\bar{v}_{\overline{A},n}\rangle_{\rm aux}\right),
\label{eq:H} 
\end{equation}
One can apply translation on both physical and auxiliary subsystems or only on physical subsystem, as shown in the schematic diagram Fig.~\ref{fig:tensor_diagram}, which yields
\begin{align}
T_{y}^\mathrm{aux} \, T_{y}^\mathrm{phy} \, \vert u_{A,j} \rangle & = e^{i\,k^p_{j} a +i \, k^{\rm a}_{j} a } \, \vert u_{A,j} \rangle \\
T_{y}^{\rm phy} \, \vert u_{A,j} \rangle &= e^{i \, k^{\rm p}_{j}a} \, \vert u_{A,j} \rangle %\left( |\bar{u}_{A,n}\rangle_{\rm aux} \otimes |u_{A,m}\rangle_{\rm phy} \right).
%\label{eq:H} 
\end{align}
The first line yields a non-chiral entanglement spectrum with a momentum $k_{j}=k^{\rm p}_j+k^{\rm a}_j$, while the second line yields a chiral entanglement spectrum with a momentum $k^{\rm p}_j$. In our calculations, we can use the basis states and study spectrum within a given momentum. In particular we can constraint the momentum in the auxiliary layer to be fixed and study entanglement spectrum with respect to momentum in the chiral layer. In the following we are going to explain how to use the momentum to extract universal chiral entanglement data, hence the corresponding chiral bulk topological data.

For this spin model, the energy, correlation functions, and transfer-matrix spectra are computed with dense tensors. For the entanglement-spectrum
analysis, we additionally use an $\rm{SU}(2)$-symmetric tensor ansatz, which resolves the total-spin quantum numbers of the Schmidt states and makes the $\rm{SU}(2)_1$ level counting directly visible.

In the numerical calculations of the entanglement spectrum in Fig.~\ref{fig:ES} in the main text, we use two complementary contraction schemes. The spectra resolved by the total momentum and by the physical momentum in Figs.~\ref{fig:ES} (a) and (b) are obtained from the CTMRG fixed-point environment, where the CTMRG boundary tensors are used to build the effective transfer-matrix fixed points along the cylinder direction~\cite{poilblanc20162}. This method allows us to reach the larger width $L_y=8$.

By contrast, the auxiliary-momentum-resolved spectra in Figs.~\ref{fig:ES} (c) and (d) require an explicit projection onto a fixed auxiliary momentum sector. The corresponding projector acts around the cylinder circumference and is implemented by exact contraction. This gives a direct separation of the auxiliary-momentum sectors, but is computationally more expensive; we therefore use a smaller width $L_y=6$ for the projected spectra.

\section{Bulk-Entanglement Boundary Correspondence of $\rm{SU}(2)_1$ Chiral Spin Liquid}
Here we first review the bulk-entanglement boundary correspondence in the $\rm{SU}(2)_1$ chiral topological order~\cite{wen_chiral_1990,wen_chiral_1991,elitzur1989remarks,li2008entanglement,qi_ent_top_2012,wen2017colloquium} and then discuss its realization in our decoupled bi-layered constructions. 

Topological point like excitations in the chiral $\rm{SU}(2)_1$ chiral topological order is described by a $\rm{SU}(2)_1$ unitary modular tensor category (UMTC). There are two simple anyon (simple objects in the tensor category language) sectors, the vacuum $I$ and semion $s$. Their fusion rules make a $Z_2$ group structure, in which the nontrivial one is
\begin{equation}
    s \times s = I.
\end{equation}
Both anyons are abelian with a quantum dimension $d_{I} = d_{s} = 1$. The semion $s$ has a non-trivial topological twist $\theta_s = i$, or equivalently a topological spin $h_s = 1/4$. As a chiral topological order, it has a chiral central charge $c_{-} = 1 \, \mathrm{mod} \, 8$. A convenient way to identify the topological data is to through its anomalous boundary.

\parR{Bulk-entanglement boundary correspondence in chiral topological order--} When putting the chiral theory on a manifold with an open boundary, the boundary has to be a gapless chiral 1+1$d$ theory -- usually a chiral CFT with a Hamiltonian~\cite{di1997conformal,blumenhagen2012introduction}
\begin{equation}
    H_{\mathrm{En}} \sim \frac{2\pi}{W} \, \left( L_0 - c/24 \right)
\end{equation}
where $W$ is the length scale of the boundary, and $L_0$ is chiral Virasoro generator. Interestingly, for a chiral CFT, the Hamiltonian and momentum denotes actually the same operator $H_{\rm En} \sim P_{\rm En}$ as a result of the global Lorentz invariance. For the opposite chirality, it becomes $H_{\rm En} \sim -\,P_{\rm En}$. %As a result, one can use the momentum to select the Hilbert space.
As a result, all primary sectors make a single right/left moving linear dispersion relation between the momentum and energy spectrum. With a given momentum, one can study spectrum at this fixed momentum and identify the primary and descendant level from the degeneracy. Note that on the lattice with a finite length scale $W$, only those low-energy states with a momentum $k \sim 0$ (or equivalently $n \ll W$ in $k=2\pi n/W$) satisfy the universal CFT prediction.

Here for the $\rm{SU}(2)_1$ chiral topological order, the universal boundary theory is a chiral $\rm{SU}(2)_1$ CFT with a central charge $c=1$~\cite{di1997conformal,blumenhagen2012introduction}. The theory can be constructed from a current operator algebra. It can also be understood as a special compact boson CFT with a particular radius. This chiral CFT has two primary sectors with a spin $0$ and $1/2$, and they are in one-to-one correspondence to anyon sectors $I$ and $s$ of the bulk theory, respectively. The boundary Hilbert space in each sector is a irreducible module labeled by a primary and its descendant states. The degeneracies within the corresponding conformal towers are encoded by the characters
\begin{equation}
\begin{aligned}
\chi_I(q)
&=
q^{-1/24}
\left(
1+3q+4q^2+7q^3+13q^4+\cdots
\right),
\\
\chi_s(q)
&=
q^{5/24}
\left(
2+2q+6q^2+8q^3+14q^4+\cdots
\right),
\end{aligned}
\label{eq}
\end{equation}
where $q=e^{2\pi i\tau}$ is the modular parameter and the character
$\chi_a(q) = \operatorname{Tr}_{\mathcal H_a}
\left ( q^{L_0-c/24}\right).$
The coefficients in Eq.~\eqref{eq} give the total level degeneracies. %including the ($\mathrm{SU}(2)$) multiplet degeneracies.

This theory also admits a global $\mathrm{SU}(2)$ symmetry -- a sub-algebra of the current operator algebra commutes with $L_0$. Then the ground state of each primary has a $\rm{SU}(2)$ spin. In particular, $s = 0$ or $s=1/2$ for the $I$ or $s$ primary respectively. Since the ladder operator generating the descendant states transform as a spin-1 object, states in each conformal tower have spin selection rule 
\begin{equation}
    j \otimes 1 = j-1 \, \oplus j \, \oplus \, j+1
\end{equation}
where spin $j-1$ disappears if $j-1<0$. As a result, the $I$ and $s$ sector contains only integer and half-integer spins, hence there is a unique $Z_2$ parity charge for each conformal tower.  

Here, in our study we consider a single cut entanglement boundary. Consider the $\rm{SU}(2)_1$ chiral topological order on a finite width $W$ and infinite length cylinder, the bipartite entanglement Hamiltonian $H_E$ defined by the half-infinite subsystem 
\begin{equation}
\rho_A = \frac{e^{-H_E}}{\mathrm{Tr}\, e^{-H_E}}    
\end{equation}
realizes this chiral CFT in the low level entanglement energies. One can use the normalized low-lying entanglement energy level counting as a fingerprint to detect the bulk $\mathrm{SU}(2)_1$ topological order.

\parR{Identifying Chiral Data from the decoupled bilayer model--} We next consider our decoupled chiral/anti-chiral bilayer construction. As a whole system, it can be formally regarded as a non-chiral one. The whole system now contains four topological sectors
\begin{equation}
        \left(\bar{I}\otimes I\right) \ \ 
        \left(\bar{I}\otimes s\right) \ \ 
        \left(\bar{s}\otimes I\right) \ \ 
        \left(\bar{s}\otimes s\right)
\end{equation}
where $\left(\bar{i} \otimes j\right)$ denotes a pair of anti-chiral/chiral sector. It admits gapped boundary conditions in principle. But here we only consider critical boundary theory, since the entanglement effect only introduces weak couplings between the two layers. Hence, the low-energy entanglement boundary contains a left-moving and a right-moving chiral critical theory. The universal boundary theory is therefore a full CFT with a Hamiltonian and momentum 
\begin{equation}
\begin{aligned}
    H_{\mathrm{En}} &\sim \frac{2\pi}{W} \left( L_0 + \overline{L}_0 - c/12 \right) \\
    P_{\mathrm{En}} &\sim \frac{2\pi}{W} \left( L_0 - \overline{L}_0\right)
\end{aligned}
\label{eq_ham_m}
\end{equation}
which has a vanishing net chiral central charge $c-=c_L-c_R=0$. The partition function on the torus has the form 
\begin{equation}
    Z_{\mathrm{full}} = \sum_{i,j}\chi_i(q) \, M_{ij} \, \overline{\chi}_{\bar{j}}(\bar{q})
\end{equation}
where $M_{ij}$ is the gluing matrix between the chiral and anti-chiral sectors, and its non-zero elements determines the Hilbert space structure. Each bulk topological sector $\left(i,j\right)$ correspondences a pair of non-zero left/right sectors, namely a non-zero $M_{ij}$, in the boundary theory. 

To identify the universal CFT data and identify the sector $\left(i,j\right)$ in the partition function, one can calculate the momentum (conformal spin) and energy Eq.~\ref{eq_ham_m}. In each sector $\left(i,j\right)$, the momentum and energy take the form of
\begin{equation}
    \begin{aligned}
        &q \sim h_i + N_i - h_{\bar{j}} - N_{\bar{j}} \\
        &\Delta \sim h_i + N_i + h_{\bar{j}} + N_{\bar{j}}
    \end{aligned}
\end{equation}
with $h_i$ the chiral conformal weight and $N_i$ the level of descendant. The degeneracy of of level of state can be obtained from the character expansion. 

However, the independent chiral and anti-chiral conformal symmetry in the boundary theory can not prove the decoupling between the chiral and anti-chiral topological order in the bulk since a general non-chiral topological order (the chiral and anti-chiral parts can not be separated on the lattice) may have the same boundary theory. Our setup is not a normal non-chiral topological order.%, such as a quantum double model, in which the chiral and anti-chiral parts can not be separated on the lattice. 
%Here we can independently manipulate the physical (chiral) and auxilliary (anti-chiral) theory to detect topological data of the chiral physical system. 
The effective couplings between the physical layer and the auxiliary layer in the wave function induced by the finite entanglement effect can affect the higher boundary entanglement levels, but they do not modify the universal low-energy organization. Put it another way, the small bulk effective couplings correspond to irrelevant deformations for the critical boundary theory. As a result, we can independently identify and control the universal low-energy part of the chiral or anti-chiral entanglement boundary theory through the physical or auxiliary bulk. 

Our PEPS wave function contains explicitly the physical and auxiliary physical index, allowing us to perform translation operator on each layers independently, as shown in Fig.~\ref{fig:tensor_diagram}. This means we can approximately label the subsystem Schmidt eigenvectors of the two layers with an independent momentum. On the chiral theory level, as explained above, the momentum operator and the entanglement Hamiltonian are actually the same one up to non-universal normalization factors. Now we can construct a fixed momentum value $k_0^a$ for the auxiliary part, and calculate the entanglement energy spectrum with respect to the physical layer momentum. The energy spectrum now has the structure
\begin{equation}
    \begin{aligned}
        &q \sim h_i + N_i - k_0^a \\
        &\Delta \sim h_i + N_i + k_0^a.
    \end{aligned}
\end{equation}
This is the universal chiral energy spectrum $\Delta \sim q$ one expects for the boundary theory of a chiral bulk topological order. 

In particular for the $\rm{SU}(2)_1$ theory, the chiral and anti-chiral global $\rm{SU}(2)$ symmetry admits a diagonal global $\rm{SU}(2)$. We consider only integer spins, hence there are $\left(I,I\right)$ and $\left(s,s\right)$ sectors in the boundary theory, while the $\left(I,s\right)$ and $\left(s,I\right)$ support half-integer spins. This parity is determined by the global physical $\rm{SU}(2)$ spin and we choose the integer spin Schmidt space, hence the boundary is limited to even parity. As shown in the main text, after choosing a zero momentum in the auxiliary layer, we indeed obtain a single linear chiral energy dispersion.

\begin{table}[htb!]
\centering
\begin{tabular}{|c|c|c|c|c|}
\hline
Sector/branch & level 0 & level 1 & level 2 & level 3 \\
\hline
\(I^{\rm phy}\)
& \(0\)
& \(1\)
& \(0\oplus 1\)
& \(0\oplus 2\times 1\) \\
\hline
\(s^{\rm phy}\)
& \(\frac{1}{2}\)
& \(\frac{1}{2}\)
& \(\frac{1}{2}\oplus\frac{3}{2}\)
& -- \\
\hline
\multicolumn{5}{|c|}{Auxiliary-momentum-projected branches} \\
\hline
\(0^{\rm aux}\otimes I^{\rm phy}\)
& \(0\)
& \(1\)
& \(0\oplus 1\)
& \(0\oplus 2\times 1\) \\
\hline
\(\frac{1}{2}^{\rm aux}\otimes s^{\rm phy}\)
& \(0\oplus 1\)
& \(0\oplus 1\)
& \(0\oplus 2\times 1\oplus 2\)
& -- \\
\hline
\end{tabular}
\caption{
Low-lying \(\rm{SU}(2)\) multiplet content of the chiral \(\rm{SU}(2)_1\) edge theory and the auxiliary-momentum-projected entanglement spectra.
The first two rows give the theoretical predictions for the physical \(I^{\rm phy}\) and \(s^{\rm phy}\) towers, while the last two rows give
the corresponding numerical results. All low-lying levels that can be unambiguously resolved in the numerical spectra are included: four levels
for the \(0^{\rm aux}\otimes I^{\rm phy}\) branch and three levels for the \(\frac12^{\rm aux}\otimes s^{\rm phy}\) branch. Within this resolved
range, the numerical multiplet content agrees level by level with the \(\rm{SU}(2)_1\) prediction. A prefactor denotes the multiplicity of the
corresponding spin multiplet. The dash indicates that the next level is not reliably resolved numerically.
}
\label{tab:su2-counting}
\end{table}

Table~\ref{tab:su2-counting} summarizes the low-lying $\rm{SU}(2)$ multiplet content used to identify the projected chiral towers in Fig.~\ref{fig:ES}.
The first two rows give the expected $I^{\rm phy}$ and $s^{\rm phy}$ towers of the chiral $\rm{SU}(2)_1$ edge theory, while the last two rows give
the auxiliary-momentum-projected branches obtained from the enlarged PEPS. The $0^{\rm aux}\otimes I^{\rm phy}$ branch directly reproduces the
identity tower. The $\frac12^{\rm aux}\otimes s^{\rm phy}$ branch is obtained by tensoring the physical semion tower with the lowest auxiliary
spin-$\frac12$ multiplet, yielding the integer-spin content shown in the table.
Importantly, the table contains all low-lying levels that can be unambiguously identified in the projected numerical spectra. We resolve
four conformal levels in the \(0^{\rm aux}\otimes I^{\rm phy}\) branch and three in the \(\frac12^{\rm aux}\otimes s^{\rm phy}\) branch, and
all of them reproduce the corresponding theoretical multiplet content.

\end{document}